\documentclass[11pt]{aip-book}
\usepackage[utf8]{inputenc}
\usepackage[T1]{fontenc}
\UseRawInputEncoding

\usepackage{graphicx,hologo,mwe,url} 
\usepackage{array,booktabs,threeparttable} 

\usepackage{amsmath}

\usepackage{amsmath}
\usepackage{rotating}

\usepackage{natbib}
\usepackage{graphicx}
\usepackage{subcaption}
\usepackage{longtable}
\usepackage{ragged2e}

\usepackage{setspace}
\usepackage{times}

\usepackage{titlesec}
\titleformat{\section}{\normalfont\Large\bfseries}{\thesection}{1em}{}
\renewcommand{\thesection}{\thechapter.\arabic{section}}

\usepackage{chngcntr}

\counterwithin{figure}{section}
\counterwithin{table}{section}
\usepackage{parskip} 

\begin{document}
\setchapter{1} 

\chapter{AI-based Approach in Early Warning Systems: Focus on Emergency Communication Ecosystem and Citizen Participation in Nordic Countries} 

\chapterauthor{Fuzel Ahamed Shaik}{Center for Machine Vision and Signal Processing}{University of Oulu, \\Finland}{fuzel.shaik@oulu.fi}
\chapterauthor{Getnet Demil}{Center for Machine Vision and Signal Processing}{University of Oulu, \\Finland}{getnet.demil@oulu.fi}
\chapterauthor{Mourad Oussalah}{Center for Machine Vision and Signal Processing}{University of Oulu, \\Finland}{mourad.oussalah@oulu.fi}

\begin{abstract}
Climate change and natural disasters are recognized as worldwide challenges requiring complex and efficient ecosystems to deal with social, economical and environmental effects. This chapter advocates a holistic approach, distinguishing preparedness, emergency responses and post-crisis phases. The role of Early Warning System (EWS), Risk modeling and mitigation measures are particularly emphasized. The chapter reviews the various Artificial Intelligence (AI)-enabler technologies that can be leveraged at each phase, focusing on INFORM risk framework and EWSs. Emergency communication and psychological risk perception have been emphasized in emergency response times. Finally a set of case studies from Nordic counties has been highlighted. 
\end{abstract}

\section{INTRODUCTION}
Climate change is a complex and multifaceted global phenomenon, characterized by long-term alterations in temperature, precipitation patterns, sea-level rise, and the increased frequency and intensity of extreme weather events. These changes are driven by anthropogenic factors, such as greenhouse gas emissions, deforestation, and industrial activities, which significantly alter the Earth's natural climate systems and render the occurrence of natural disasters inevitable. Climate-related catastrophes, such as hurricanes, floods, droughts, wildfires, heatwaves, and rising sea levels, have become increasingly frequent and severe in recent years, affecting billions of people globally, and this trend is expected to continue in the future. Indeed, the Emergency Events Database (EM-DAT) estimates that between 3.3 to 3.6 billion people are exposed to extreme risk as a result of climate-related disasters \citep{keim2021}. Natural disasters alone impact approximately 200 million people annually, as reported by the United Nations (UN) \citep{dwivedi2022}. Despite major investments in advanced early warning systems (EWSs) to lessen the effects of these natural catastrophes, there still needs to be more public awareness, effective interaction with various communities, and accurate prediction to minimize societal, economic, and environmental damage. This emphasizes the critical necessity for effective crisis communication tactics to reduce the effects of these disasters and allow active user participation in disaster prevention and restoration efforts. In its basic concept, an EWS is an infrastructure that yields a set of signals that facilitate emergency communication between a minimum of two social groups: the alerting authority and the directly affected population \citep{fischer2020}. For authorities, the effective use of EWS involves verified and identified communications that prompt timely countermeasures by the population, while from the population perspective, it subsumes receiving actionable information to manage ongoing emergencies, thus enabling self-protection \citep{fisher2001}.

Information and Communication Technology (ICT) plays a crucial role in the design of EWS infrastructure \citep{shaik2024} by providing vital insights into aggregating sensory data for monitoring hazards, such as active volcanoes, rising sea levels, cyclone propagation, as well as enhancing both the prediction capabilities on the impacts of climate change effects and interaction with various end-users. For example, sensor systems, such as ocean buoys and seismic sensors, enable real-time data collection and analysis via satellites, supported by grid computing for comprehensive insights \citep{esposito2022}. Similarly, the Global Earth Observation System provides universal access to crucial data, enhanced by GIS for visualizing environmental data crucial for climate change initiatives and ecosystem protection. Climate solution research can nowadays design policies and implement them at building, street, household, and urban region levels, and these policies can be tailored to unique situations while still being scalable to global mitigation opportunities, thanks to the development of big data and Artificial Intelligence (AI) / Machine Learning (ML) tools \citep{dwivedi2022}. In this context, AI and ML present encouraging prospects to strengthen citizen awareness and EWS accessibility worldwide. Although several state-of-the-art ML and deep-learning models have been put forward since the last decade, their application to climate-change prediction, analysis, and EWSs has been limited until recently. Indeed, this has been possible thanks to recent advances in high-performance computing, cloud computing, efficient IoT, and sensory array systems where several prototypes have been put forward. \citep{haggag2021deep} advocated a deep learning model, which analyzes spatiotemporal climate data, for predicting climate-induced disasters, showcasing how advanced algorithms can significantly enhance the accuracy and reliability of disaster forecasts. Famine Early Warning System Network (FEWSN) utilizes ICT, advanced classification model, and scenario development process to predict famine by analyzing climate and weather impacts on crops, aiding decision-makers with monthly updates and alerts for emergency preparedness \citep{dwivedi2022}.  

In addition to FEWSN, other advanced Early Warning Systems (EWS) have been developed to address various environmental hazards. The Global Flood Awareness System (GloFAS), a joint initiative of the European Commission and the European Centre for Medium-Range Weather Forecasts, provides global flood forecasts and early warnings. GloFAS utilizes advanced hydrological and meteorological models, satellite data, ground observations, and machine learning algorithms to generate daily forecasts up to 30 days in advance, aiding decision-makers in flood risk management and emergency preparedness \citep{emerton2016glofas}. Similarly, the European Forest Fire Information System (EFFIS) offers comprehensive information for forest fire protection in Europe and neighboring regions. EFFIS employs remote sensing data, meteorological forecasts, and advanced modeling techniques to assess fire danger, predict fire behavior, and provide daily fire danger forecasts and monitoring. This system supports authorities in fire prevention, preparedness, and post-fire damage assessments \citep{sanmiguelayanz2012effis}. Both GloFAS and EFFIS, like FEWSN, demonstrate the integration of Information and Communication Technology (ICT), advanced modeling, and data analysis techniques, including artificial intelligence, to provide early warnings for different types of hazards, supporting decision-makers with timely and actionable information for emergency preparedness and response phases. 

Overall, artificial intelligence contributes to EWS from different standpoints. First, from a sensory data perspective, it enables efficient real-time measurements from disparate sources that include satellites, sensor arrays, IoT, and complex multi-modality observation systems installed at meteorological centers and earth science observation sites. Second, it is continuously being employed in the modeling and predictive tools where deep learning models have achieved state-of-the-art performances in several predictions related to climate change effects and disaster monitoring. Third, it enables efficient interaction with users and various stakeholder groups to maximize users' awareness and contribution towards the mobilization effort. This is achieved, for instance, through third-party apps that intelligently exploit users' profiles and interests using tailored recommender systems and emergency apps. This collectively enhances both community and individual resilience to climate change and natural disaster effects.     

This chapter aims to detail the above holistic view by summarizing and exemplifying the contribution of AI in the different phases of climate change and/or natural disaster prediction, monitoring, and management, highlighting its association with EWSs, risk analysis, and emergency communications as part of global human-machine interaction schemes, for efficient management and resilience build up. For the risk analysis part, the chapter focuses on the EU-accredited INFORM risk framework \citep{de2015} and examines how AI-based reasoning and technologies can be encapsulated in various components of the INFORM risk framework. Finally, a set of selected case studies from Nordic countries are examined in this context. Section 2 summarizes worldwide initiatives in tackling climate change and natural disaster management that impacted the contribution of AI technology in this field. Section 3 details how AI can be encapsulated in various components of the INFORM risk framework. Section 4 emphasizes emergency communications and the way AI influences users' responses. Section 5 focuses on selected Nordic case studies, while Section 6 draws summarizes the key findings and perspective work.  \label{sec:1}
\section{\MakeUppercase{Holistic view in disaster management}} 
\begin{sidewaysfigure}
    \centering
    \includegraphics[width=0.95\linewidth]{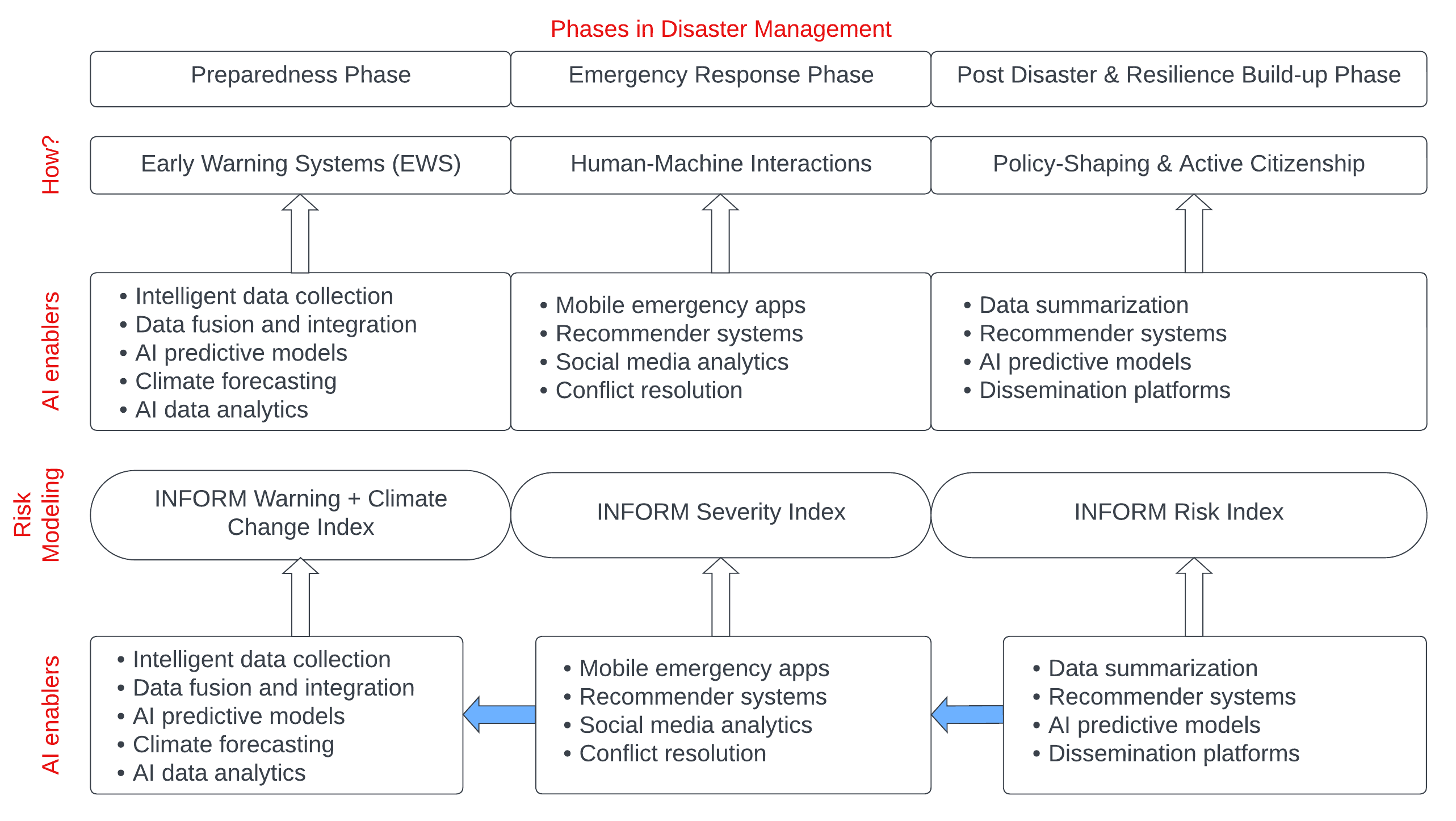}
    \caption{Overview illustration of AI enablers in disaster preparedness, emergency response, and post-disaster phases.}
    \label{fig:overview}
\end{sidewaysfigure}

The increasing frequency and severity of climate-related disasters call for comprehensive global initiatives leveraging comprehensive theoretical frameworks and robust early warning systems. These systems not only predict and monitor potential disasters but also disseminate warnings effectively to ensure timely response and mitigate impacts, motivating a holistic approach towards natural disaster prediction and management. In this respect, the INFORM risk framework, and Sendai Framework, among others, e.g., \cite{papathoma2016common}, are key global initiatives that combine sound theoretical framework, fine-grained sensing technologies embedded in EWSs, and evidence-based interventions. Especially, the INFORM framework uses a set of performance indicators to measure, monitor, and manage risks associated with disasters at its various phases (preparedness, emergency response, and post-disaster), see Fig. \ref{fig:overview}. 

Overall, EWS plays a vital role in mitigating climate-related disasters by predicting extreme weather events and issuing timely alerts, allowing authorities and the public to take proactive measures. While technologies, like emergency mobile communications, geo-localization, object recognition and tracking from UAV and satellite data, social interaction, and recommendations play a key role in emergency response phase, where locating people at risk, identifying damages, and ensuring effective communication and collaboration of various stakeholders, including citizens, is of paramount importance. Integrating AI into EWS enhances their effectiveness by improving the accuracy and speed of processing complex climate data, thus revolutionizing the data collection process, risk assessment, communication, and personalized recommendations. The ultimate goal of risk frameworks and EWSs is to enhance community resilience. For instance,  \citep{abrash2021building} reviewed effective community resilience interventions in the Northeastern United States, identifying strategies like the COAST project and Ready CDC intervention. These initiatives enhance public health outcomes by improving emergency preparedness and mental health resilience. 

In essence, community resilience is attained through a combination of policing, social connectedness, and tailored interventions. Regulatory frameworks enforce acknowledged climate-change practices in construction, urban planning, and environment protection guidelines. Social connectedness is enforced through efficient communication strategies that maintain social cohesion and sustain trust in mass media and national organizations. Social media, emergency communications, education, and various dissemination portals contribute to this goal. Finally, tailored interventions capitalize on distinguished characteristics of communities or individuals to devise targeted interventions through recommender systems, business incentives, and other instruments to tackle specific needs identified at the community or individual level. These studies highlight the need for standardized tools to measure and enhance community resilience, emphasizing its alignment with national preparedness goals. This aspect is mainly tackled through local and national policy documents where the role of various actors and key objectives together with the associated interventions are devised.



\label{sec:2}
\section{\MakeUppercase{Mapping climate-change related disasters and worldwide initiatives}}

\subsection{Overview of Global Disasters}
The global landscape of natural disasters presents a complex interplay of various natural and human-induced factors, highlighting the critical need for comprehensive disaster risk management strategies. Geographical factors play a significant role in determining the nature and severity of these hazards. For instance, the 2008 Wenchuan earthquake in China resulted in a cascade of secondary disasters such as rock falls, landslides, and debris flows, while the 2011 Eastern Japan earthquake triggered a tsunami and a subsequent nuclear disaster \citep{norio2011, zhou2015}. These contrasting events underscore how different geographical, urbanization and industrial settings can influence the progression and impact of disasters. Recent advances in disaster risk science have significantly enhanced our understanding of natural hazards and improved our ability to forecast and mitigate their impacts. Research has expanded across various disciplines, including seismology, meteorology, geology, hydrology, and geography, to comprehend the formation mechanisms and dynamics of natural hazards like earthquakes, tropical cyclones, floods, and wildfires \citep{emanuel2017assessing}. 
This interdisciplinary approach has been instrumental in developing more effective early warning systems and improving our overall preparedness for these events \citep{schiermeier2018climate, mazdiyasni2015substantial, ward2020impact}. On a global scale, the impact of natural hazards is profound. Since 1990, natural disasters have caused over 1.6 million fatalities and resulted in economic losses averaging between USD 260-310 billion annually. The international community has recognized the need to mitigate these risks through frameworks like the Sendai Framework for Disaster Risk Reduction, which highlights the necessity of integrating disaster risk reduction into sustainable development goals \citep{center2015sendai} and the INFORM risk framework.

\begin{figure}
    \centering
    \includegraphics[width=0.75\linewidth]{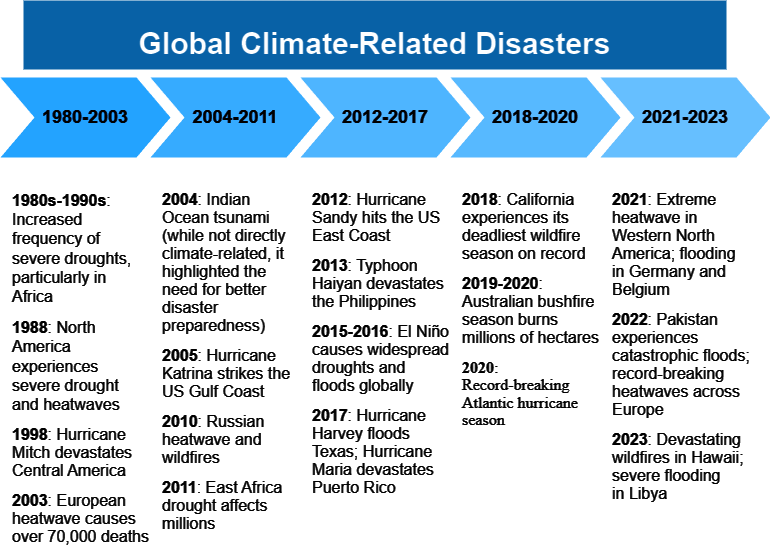}
    \caption{Timeline of Climate-Related Disasters}
    \label{fig:timeline}
\end{figure}

\subsection{Global Initiatives}
\begin{figure}
    \centering
    \includegraphics[width=1\linewidth]{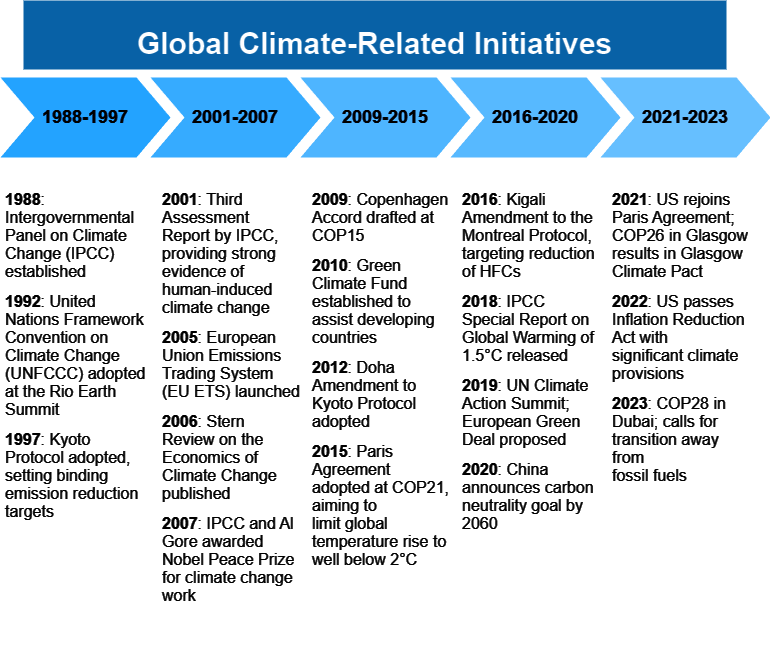}
    \caption{Timeline of Climate-Related Global Initiatives}
    \label{fig:timeline2}
\end{figure}

Worldwide and national initiatives have been launched to combat climate change and enhance resilience in response to these escalating disasters. Governmental grassroots movements and non-governmental organizations are playing crucial roles in raising awareness and driving local action. Collectively, these initiatives reflect a global commitment to addressing the multifaceted challenges posed by climate change-related natural disasters \citep{church2023}. ICT and AI have emerged as powerful tools in the global fight against climate change as well. Various international organizations, governments, and private sector entities have recognized their potential and are actively incorporating these technologies into their climate strategies. A timeline of key recent climate related disaster events and initiatives programs taken from the authorities are summarized in Figures \ref{fig:timeline} and \ref{fig:timeline2}, respectively. Besides, a selection of these initiatives is provided below.

\textbf{Paris Agreement and Technology Framework} \citep{unfccc2015paris}, adopted in 2015, explicitly acknowledges the critical role of technology in addressing climate change. The implementation plan involves strengthening cooperative action on technology development and transfer, including through the Technology Mechanism and the Financial Mechanism of the Convention. Countries are encouraged to include ICT and innovative technologies in their Nationally Determined Contributions (NDCs) and long-term strategies.


\textbf{European Green Deal and Digital Strategy} \citep{com2019european}, launched in 2019, places significant emphasis on digital technologies as key enablers for achieving climate objectives. The European Commission's digital strategy, "Shaping Europe's Digital Future," outlines how digital technologies, including AI, can accelerate and maximize the impact of policies to deal with climate change and protect the environment. 


\textbf{IPCC 6th Assessment Report and AI in Climate Modeling} \citep{masson2021climate}, released in 2021, acknowledges the growing role of AI in climate science. The report highlights how machine learning and AI techniques are enhancing climate modeling, improving the accuracy of climate predictions, and supporting impact assessments. 

\textbf{United Nations Environment Programme (UNEP) and AI for Earth Monitoring}, has been at the forefront of leveraging AI for environmental monitoring and climate action. In its 2020 report \citep{unep2020frontier}, UNEP outlines how AI is being used for: i) Analyzing satellite imagery to track deforestation, land use changes, and biodiversity loss; Optimizing renewable energy systems; iii) Improving climate change predictions and impact assessments.

\textbf{World Economic Forum: AI for Climate Action} \citep{wef2018ai}, provides a comprehensive overview of AI applications in climate change mitigation and adaptation. The report outlines six priority areas for AI application: Climate change mitigation and adaptation, Biodiversity and conservation, Healthy oceans, Water security, Clean air, and Weather and disaster resilience. 


\textbf{Microsoft AI for Earth} \citep{smith2017ai}, launched in 2017, represents a significant private sector initiative in this space. The program provides cloud and AI tools to organizations working on sustainability and climate issues. The implementation strategy includes: Providing grants of Azure cloud computing resources, Offering AI tools and training to researchers and organizations, and Supporting the development of open-source tools for environmental applications. 

\label{sec:3}
\section{\MakeUppercase{AI in Risk Management (INFORM)}}

Launched in 2014, the INFORM Risk Index provides a unified framework and methodology for evaluating humanitarian risk based on structural factors \citep{de2015}. INFORM's products enhance disaster risk management through four key components namely, \textit{INFORM Risk} for crisis prevention and preparedness, \textit{INFORM Warning} for early action and emerging crisis detection, \textit{INFORM Severity} for assessing ongoing crisis severity, and \textit{INFORM Climate Change} for addressing risk reduction and climate adaptation. These products collectively improve crisis response and adaptation strategies, leveraging transparent, data-driven methods and years of accumulated expertise \citep{poljanvsek2020, benini2016}.

\subsection{INFORM Risk}
According to the INFORM report (2024) \citep{inform2024}, the INFORM risk index represents a pioneering global tool that offers an objective and transparent overview of the risk of humanitarian crises and disasters. The index's primary goal is to identify potential crisis hotspots, allowing for proactive measures to reduce risk, build resilience, and prepare for potential crises. Its insights enable us to pinpoint vulnerable areas, understand the underlying causes, and take proactive measures to reduce risk, bolster resilience, and improve preparedness for potential crises. It employs 80 different indicators to assess a wide range of hazards, people's exposure to these hazards, their vulnerability, and the availability of resources to help them cope with these risks. This index thoroughly evaluates risk factors, allowing for a more nuanced understanding of potential threats and effective response strategies. The index is structured around three main dimensions: Hazard and Exposure, Vulnerability, and Lack of Coping Capacity. These dimensions are further broken down into categories and components, creating a comprehensive framework for risk assessment. 


AI and ICT can significantly enhance the INFORM Risk Index's capabilities, making it more dynamic, accurate, and actionable in the following way. 
\begin{itemize}
    \item \textbf{Advanced Data Collection and Integration}:
AI-powered data collection systems can significantly enhance the INFORM Risk Index's data gathering process. Natural Language Processing (NLP) algorithms can be employed to continuously scan and extract relevant information from diverse sources such as news articles, social media, and scientific reports \citep{imran2015processing}. This real-time data collection can be particularly valuable for capturing rapidly evolving risks. Machine learning techniques, particularly deep learning models like transformers, can be used to process and integrate unstructured data, converting it into structured formats compatible with the INFORM Risk framework \citep{devlin2018bert}. This could allow the inclusion of more diverse data sources, potentially uncovering new risk factors or improving the accuracy of existing ones. Moreover, federated learning techniques could enable data sharing and model training across multiple organizations without compromising data privacy, facilitating more comprehensive risk assessments \citep{yang2019federated}.

    \item \textbf{Intelligent Risk Modeling}:
The INFORM Risk Index could benefit from advanced machine learning models for risk assessment. Ensemble methods combining multiple AI algorithms (e.g., random forests, gradient boosting machines, and neural networks) can be employed to capture complex, non-linear relationships between risk factors \citep{sagi2018ensemble}. These models can be trained on historical data to identify subtle patterns and interactions that might be missed by traditional statistical approaches. Deep learning architectures, such as Long Short-Term Memory (LSTM) networks, can be utilized to model temporal dependencies in risk factors, capturing how risks evolve over time \citep{hochreiter1997long}. This could provide a more dynamic understanding of risk trajectories.

    \item \textbf{Dynamic and Adaptive Risk Assessment}:
AI can transform the INFORM Risk Index from a static annual assessment into a dynamic, continuously updated system. For instance, reinforcement learning algorithms can be employed to adapt the risk assessment model in real-time as new data becomes available \citep{arulkumaran2017deep}. This approach would allow the index to quickly respond to emerging risks or sudden changes in the global risk landscape. Online learning algorithms can be integrated to continuously refine the model's predictions based on the latest data, ensuring that the risk assessments remain current and relevant \citep{hoi2018online}.

    \item \textbf{Advanced Predictive Analytics}:
AI-driven predictive models can significantly enhance the INFORM Risk Index's forecasting capabilities. Bayesian networks and probabilistic graphical models can be used to model the causal relationships between different risk factors, allowing for more nuanced predictions of how changes in one area might impact overall risk \citep{pearl2009causality}. Time series forecasting models, such as Prophet or ARIMA, enhanced with machine learning techniques, can project risk trends into the future, accounting for seasonal patterns and long-term trends \citep{taylor2018forecasting}. These projections could help organizations anticipate future crises and plan preventive measures.

    \item \textbf{Intelligent Visualization and Reporting}:
Advanced ICT solutions can transform how INFORM Risk data is visualized and reported. Interactive, AI-powered dashboards using technologies like D3.js or Tableau can provide dynamic, user-friendly interfaces for exploring risk data \citep{bostock2011d3}. These dashboards can incorporate machine learning algorithms to suggest relevant visualizations based on the user's role and interests. Natural Language Generation (NLG) techniques can be employed to automatically generate narrative reports explaining risk assessments in plain language, making the insights more accessible to non-technical stakeholders \citep{gatt2018survey}.

    \item \textbf{Enhanced Early Warning Systems}:
By integrating AI with the INFORM Risk Index, more sophisticated early warning systems can be developed. Anomaly detection algorithms, such as Isolation Forests or Autoencoders, can be used to identify unusual patterns in risk data that might indicate an impending crisis \citep{chalapathy2019deep}. Graph neural networks can be employed to model the complex interconnections between different risk factors and geographical regions, potentially uncovering cascading risks that might be missed by simpler models \citep{wu2020comprehensive}.

    \item \textbf{Advanced Scenario Planning and Simulation}:
AI can power complex, agent-based simulations using INFORM Risk data. These simulations can model the behavior of multiple actors (e.g., governments, NGOs, affected populations) in crisis scenarios, providing insights into potential outcomes and the effectiveness of different interventions \citep{bonabeau2002agent}. Generative AI models, such as Generative Adversarial Networks (GANs), could be used to generate synthetic crisis scenarios, helping organizations prepare for a wider range of potential futures \citep{goodfellow2014generative}.

    \item \textbf{Contextual and Adaptive Localization}:
Transfer learning techniques can be employed to adapt the global INFORM Risk model to specific local contexts. This approach allows the model to leverage knowledge gained from data-rich regions while still capturing unique local risk factors \citep{pan2009survey}. Hierarchical Bayesian models can be used to simultaneously model risks at different geographical scales (global, regional, national, local), capturing both broad trends and local specificities \citep{gelman2006data}.

    \item \textbf{Automated Trend Analysis and Insight Generation}:
AI algorithms, particularly those in the field of Automated Machine Learning (AutoML), can be used to automatically identify and report on emerging risk trends. These systems can continuously analyze the INFORM Risk data, detecting significant changes, correlations, and patterns without human intervention \citep{he2021automl}. Explainable AI techniques, such as SHAP (SHapley Additive exPlanations) values, can be used to provide interpretable insights into why certain areas are assessed as high-risk, enhancing transparency and trust in the system \citep{lundberg2017unified}.

    \item \textbf{Enhanced Data Quality and Reliability}:
AI can play a crucial role in ensuring the quality and reliability of the INFORM Risk Index. Anomaly detection algorithms can flag unusual or potentially erroneous data points for human review \citep{chandola2009anomaly}. Machine learning models can be trained to estimate missing data, reducing gaps in the risk assessment. Techniques like multiple imputation or matrix factorization can be employed to handle missing data in a principled manner \citep{van2018flexible}. By incorporating these advanced AI and ICT technologies, the INFORM Risk Index can evolve into a more dynamic, accurate, and actionable tool for global risk assessment and crisis prevention.
\end{itemize}

\subsection{INFORM Warning}
According to the INFORM report (2024) \citep{inform2024}, INFORM Warning addresses the urgent need for open, aggregated, and multi-hazard early warning information that supports effective decision-making in crisis management. Extensive research and consultations with INFORM Partners have highlighted the necessity for such a system, especially as the multilateral system increasingly emphasizes anticipatory and early action. Despite the growing volume and speed of information sources available for humanitarian early warning, significant challenges persist in translating this data into actionable decisions that can reduce the frequency and severity of crises. Current early warning systems are often not open, lack global coverage, and fail to encompass all relevant risk drivers, making them both cost and time-intensive for organizations to utilize effectively. The primary goal of INFORM Warning is to bridge the gap between the INFORM Risk Index, which assesses structural crisis risk, and the INFORM Severity Index, which measures the severity of ongoing crises. 



The main elements of the INFORM warning systems are Risk monitor, Analysis, Human curation, and web platforms
\begin{itemize}
     \item \textbf{Risk Monitor}: This initial component receives risk indicators, forecasts, and warnings from trusted external sources. These inputs serve as the foundation for subsequent analysis, and collecting data related to potential crises.
    \item \textbf{Analysis}: In this stage, the system processes and normalizes the collected data. Existing predictive models generate forecasts based on risk indicators, resulting in risk assessments and predictions.
    \item \textbf{Human Curation}: Human analysts curate, prioritize, and validate the system’s outputs. Their expertise ensures accuracy and relevance, bridging the gap between automated analysis and real-world decision-making.
    \item \textbf{Web Platform}: The web-based platform presents the system’s outputs to users, allowing exploration, feedback, and informed actions based on predictions. Understanding this flowchart provides insights into disaster preparedness and response. 
\end{itemize}

Supporting INFORM Warning with ICT and AI involves leveraging advanced technologies to collect, process, analyze, and disseminate data efficiently. AI can significantly enhance the INFORM Warning system's capabilities in collecting, analyzing, aggregating, and presenting multihazard information for crisis prediction and prevention. Machine learning algorithms can be employed to process the vast amount of data from various sources, identifying patterns and trends that human analysts might miss \citep{akter2019big}. Natural Language Processing (NLP) techniques can be used to extract relevant information from unstructured text data, such as news reports and social media, to provide real-time insights into emerging risks \citep{imran2015processing}. For the quantification of risk factors, AI models can be developed to standardize and normalize data from diverse sources, enabling more accurate comparisons across different hazards and regions \citep{senaratne2017review}. 

AI-driven anomaly detection algorithms can be implemented to identify unusual patterns or sudden changes in risk indicators, alerting decision-makers to potential emerging crises \citep{chalapathy2019deep}. Furthermore, AI can support the development of interactive visualization tools that allow users to explore complex risk data intuitively, facilitating better understanding and decision-making \citep{andrienko2020visual}. To address the challenge of incorporating information into decision-making processes, AI-powered recommendation systems can be developed to suggest prioritized actions based on the analyzed risk data \citep{chen2017disease}. These systems can take into account the specific context and resources of different organizations, providing tailored recommendations for anticipatory action. Lastly, AI can contribute to the continuous improvement of the INFORM Warning system through reinforcement learning techniques, which can adapt and refine the models based on feedback and observed outcomes over time \citep{arulkumaran2017deep}. This would ensure that the system remains accurate and relevant in the face of changing global risk landscapes.

\subsection{INFORM Severity}
The INFORM Severity Index (SI) evaluates the humanitarian impact of crises, whether man-made, natural, or both, using a structured and continuously updated methodology to assist decision-makers and enhance humanitarian coordination. The SI is calculated using an analytical framework with three dimensions: the Impact of the Crisis (IC), the Condition of People Affected (CPA), and the Complexity of the Crisis (CC), which are combined as \citep{poljanvsek2020} 
\[Severity Index (SI) = IC * CPA + CC\] 
This formula underscores the interconnectedness of these factors in determining crisis severity, with each dimension further divided into components and categories based on various indicators. The INFORM SI aggregates information from credible, publicly available sources and includes ongoing monitoring of changes and new data by human analysts. The Index is updated monthly based on the availability of new, reliable information, providing a current snapshot of the humanitarian situation. Indicators are updated as soon as more accurate data is available, reflecting the most up-to-date assessments of each crisis. Despite leveraging relevant data to classify the SI of each country and monitor the change in SI every 3 months, this methodology has yet to address some limitations as mentioned in \citep{poljanvsek2020} such as Data Quality and Availability, Precision and Bias, and Data Interpretation. To make this model error-free, current ICT methods can be leveraged, some of the cases are described below:

\begin{itemize}
    \item \textbf{Satellite Imagery and Remote Sensing:}
     Satellite imagery provides real-time data on physical damages, environmental changes, and population movements. Remote sensing technology can be used to monitor disaster-affected areas continuously. During the 2010 Haiti earthquake, satellite imagery was crucial for mapping damage and coordinating rescue efforts \citep{kerle2011}. Organizations like the United Nations Satellite Centre (UNOSAT) provided detailed maps to support relief operations \citep{lang2020}.
    \item \textbf{Big Data Analytics:}
     Big data analytics can process large volumes of data from various sources, including social media, news reports, and on-ground sensors. This can help identify patterns and trends that might not be visible through traditional data collection methods. The Global Database of Events, Language, and Tone (GDELT) uses big data analytics to monitor global human society by analyzing news media from around the world in real-time \citep{wang2022}.
    \item \textbf{Mobile Data Collection:}
     Mobile technology can be used for rapid data collection in the field. Applications like KoBoToolbox \footnote{\url{www.kobotoolbox.org/}} allow humanitarian workers to collect and transmit data even in remote or offline areas. During the Ebola outbreak in West Africa, in 2014, mobile data collection tools were used to track the spread of the disease and manage response efforts efficiently \citep{sacks2015}.
    \item \textbf{Geographic Information Systems (GIS):}
     GIS tools can analyze spatial data to visualize and assess the geographic distribution of crises. GIS can integrate data from multiple sources to provide a comprehensive view of the crisis's impact. GIS was extensively used in the aftermath of Hurricane Katrina, in 2005, to map flooded areas, damaged infrastructure, and population displacement \citep{klemas2015}.
    \item \textbf{AI and Machine Learning:}
     AI and machine learning algorithms can predict the severity and spread of crises by analyzing historical data and real-time inputs. These technologies can also automate data cleaning and analysis processes. AI was used to predict the spread of COVID-19 by analyzing travel patterns and social interactions, helping to implement timely interventions.
\end{itemize}

The INFORM SI aggregates various data from social media platforms using multi-modular data fusion techniques to enhance data dependability and forecast accuracy. It provides real-time, on-the-ground data, improving situational awareness and coordination among responders. Mobile applications aid in strategy planning and resource allocation by accelerating data collection on impacted demographics and resource requirements. AI maximizes resource distribution for recovery operations, while ICT tools like GIS produce comprehensive maps of critical locations. Live satellite monitoring systems offer routing recommendations during calamities, reducing reliance on erroneous field reports. Big data analytics and AI improve data precision and reduce biases by automating intricate analysis and combining several data sources. Responders benefit from visualizing data patterns and trends using data visualization tools like Tableau, Power BI, and Looker. Deploying analytical data pipelines on cloud services speeds up data processing, improving decision-making effectiveness and enabling a robust, scalable infrastructure for emergency response.

\subsection{INFORM Climate Change}
The INFORM Risk Index has been expanded to include the INFORM Climate Change Risk Index. With the use of integrated climatic and demographic forecasts, this tool assesses the dangers that climate change poses today and in the future for a range of emission and population scenarios \citep{poljanvsek2022}. In layman's terms, it considers current and future risks by incorporating climate and demographic projections, thereby providing insights into how risks will evolve under different scenarios. By offering an evidence-based assessment of potential consequences and vulnerabilities, it is intended to support decision-making for Disaster Risk Reduction (DRR) and Climate Change Adaptation (CCA) \citep{poljanvsek2022}. The conceptual framework designed to calculate the climate change risk index is similar to the INFORM Risk Index. A multiplicative equation is utilized to determine the risk, with equal weights assigned to each axis (Hazard, Vulnerability, and Lack of Coping Capacity) at 33\% \citep{poljanvsek2022}. This guarantees that the risk assessment takes into account the internal and external factors that influence risk. Five risk categories are used to group the results to facilitate interpretation and prevent over-precision. For instance, this methodology in action could be evaluating the risk of drought in a specific country. The index would assess current drought exposure, project future drought conditions under various climate scenarios, and analyze the socio-economic vulnerability and coping capacity of the population to estimate future risk levels. This helps in identifying areas that need targeted DRR and CCA interventions. This index poses some demerits, such as oversimplifying complex realities, which can lead to simplistic policy conclusions, and the use of proxy measures may not fully capture all relevant phenomena, potentially leading to gaps in risk assessment \citep{poljanvsek2022}. 

\begin{figure}[ht]
    \centering
    \includegraphics[width=0.75\linewidth]{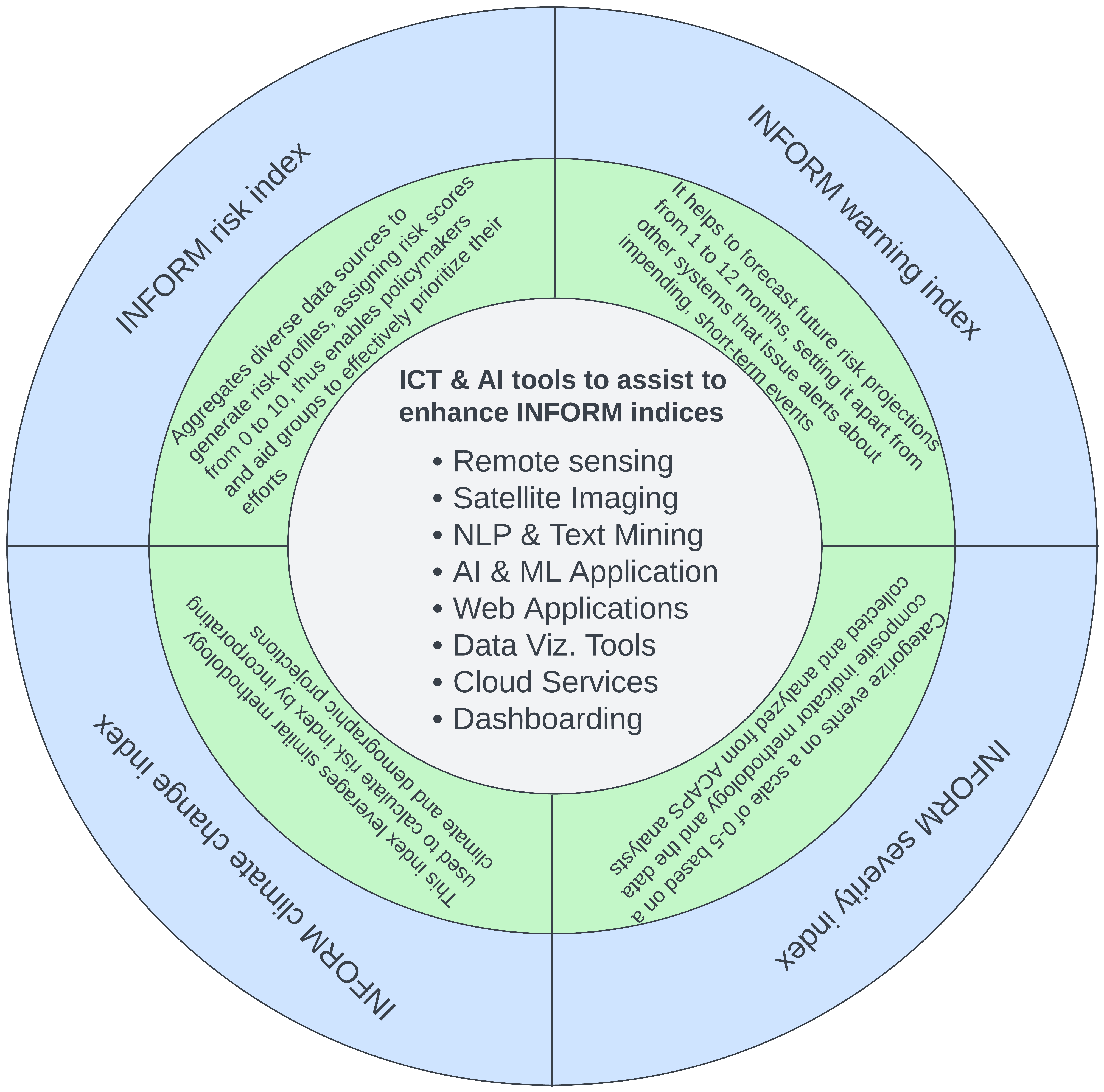}
    \caption{Integrated INFORM framework with ICT and AI tools.}
    \label{fig:inform}
\end{figure}

The efficacy of the INFORM Climate Change Risk Index can be greatly increased by including AI and ICT. By using machine learning models for precise predictive analytics, detecting subtle patterns and trends in data, and performing sophisticated data analysis, artificial intelligence can increase the precision of risk assessment. AI can also swiftly simulate a wide range of scenarios and provide detailed results for potential hazards in the future. ICT can support real-time monitoring and data collection to ensure timely and pertinent assessments. It can also enable the integration of various data sources, such as social media and satellite imaging, for a more comprehensive understanding of hazards. Moreover, interactive decision support systems can be created with ICT, enabling decision-makers to examine different risk scenarios and reach well-informed conclusions. These advancements will make the INFORM Climate Change Risk Index more dynamic, accurate, and valuable for disaster risk reduction and climate adaptation strategies. Moreover, AI-driven Natural Language Processing (NLP) can enhance the understanding and communication of risk reports, while blockchain technology can ensure data integrity and transparency in risk assessments.
\label{sec:4}
\section{\MakeUppercase{AI in emergency response phase}}
\subsection{Scope of AI-based approaches.}
\begin{figure}
    \centering
    \includegraphics[width=0.95\linewidth]{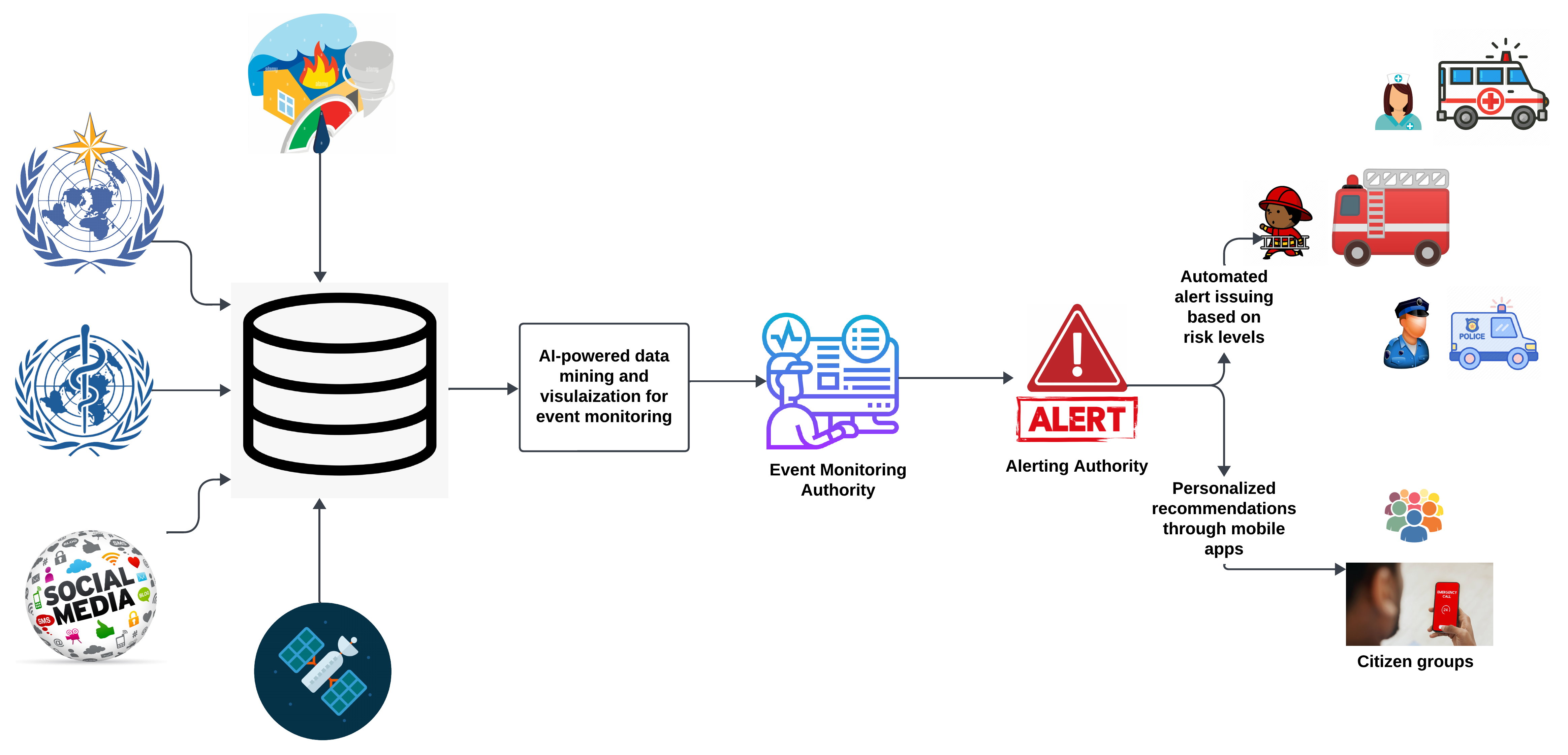}
    \caption{An illustration of the impact of AI in Risk assessment and EWS.}
    \label{fig:info}
\end{figure}

In the emergency response phase, the key is to ensure effective communication with various stakeholder groups and citizens in a way that enables effective discrimination of relevant knowledge among the vast amount of information fueled to emergency centers and ensures effective citizens' participation and response. An illustration of AI dealing with information overloading and miscommunication among various stakeholder groups in an emergency is shown in Figure \ref{fig:info}. In this course, AI systems can efficiently process and analyze vast datasets, providing timely insights for prompt decision-making. For instance, AI-driven predictive models offer accurate risk forecasts, which can be communicated to stakeholders for proactive mitigation. Automatic text summarization tools and information retrieval systems (\cite{Mohamed2019} provide enhanced capabilities for scrutinizing large amounts of textual data to identify relevant cues that can be further investigated by relevant authorities. Interactive AI-powered Decision Support Systems (DSS) integrate data from multiple sources, ensuring comprehensive and transparent communication between decision-makers and operational teams. These DSS tools centralize relevant data, making it accessible and easy to interpret, thus ensuring a shared understanding of risks and required actions. AI also enhances the visualization of complex risk data through detailed heat maps and models, making information more accessible to non-technical stakeholders. 

Moreover, AI can automate the generation of consistent and timely risk assessment reports, reducing manual workloads and ensuring regular updates for stakeholders. AI's continuous learning capability ensures that communication practices remain current with evolving risk landscapes, providing stakeholders with the most relevant information. For instance, \cite{alam2020} highlights the importance of AI techniques in enhancing situational awareness among various stakeholder groups after the 2017 Atlantic Hurricane disaster events by considering textual and image data from the microblogging platform Twitter. A significant advantage of AI in communication is the automation of risk assessment report generation. AI can compile data, analyze trends, and produce consistent, timely reports without manual intervention, reducing the workload on human analysts and ensuring regular dissemination of updated reports to stakeholders. \cite{schwarz2023} designed an AI-powered automated analytical pipeline to generate awareness reports which allowed the stakeholder groups to draw relevant insights regarding the crisis event. Automated reporting ensures that all relevant parties have access to the most current risk assessments, allowing for continuous monitoring and timely responses to emerging threats. By integrating AI into communication and dissemination processes, risk management becomes more proactive, enabling organizations to anticipate potential risks and implement mitigation strategies before they materialize. This proactive approach minimizes disaster impacts and enhances the overall resilience of communities and infrastructure.

A short assessment of ICT and AI tools and their scope of implementation in various phases of disasters are shown in Table \ref{tabai}. The next two subsections illustrate AI applications in emergency communication including mobile apps and risk perception.

\begin{longtable}{>{\RaggedRight}p{5cm} >{\centering\arraybackslash}p{1.75cm} >{\centering\arraybackslash}p{1.75cm} >{\centering\arraybackslash}p{1.5cm} >{\centering\arraybackslash}p{1.75cm} >{\centering\arraybackslash}p{1.75cm}}
  \caption{Leveraging ICT and AI tools in risk and disaster management} \label{tabai} \\
  \toprule
   & Event Monitoring & Risk Assessing & Alerting Systems & Responsive Phase & Resource Allocation \\
  \midrule
  \endfirsthead
  \multicolumn{6}{c}{{\tablename\ \thetable{} -- Continued from previous page}} \\
  \toprule
   & Event Monitoring & Risk Assessing & Alerting Systems & Responsive Phase & Resource Allocation \\
  \midrule
  \endhead
  \bottomrule
  \endfoot
  \bottomrule
  \endlastfoot
        Satellite imaging and remote sensing & X & X & X & X & X \\
        \\
        AI integrated Emergency Mobile apps & X & X & X & -- & -- \\
        \\
        Interactive AI-powered DSS  & -- & X & -- & X & X \\
        \\
        Personalized recommendation systems & -- & X & X & X & X \\
        \\
        ML-based predictive models & X & X & X & X & X \\
        \\
        Big data analytics & X & X & X & X & X \\
        \\
        Aerial hotspot mappings & X & X & X & X & -- \\
        \\
        Scenario planning and simulation & -- & X & -- & X & -- \\
        \\
        Text summarization and visualization & X & X & -- & X & -- \\
        \\
        Cloud-based implementation & X & X & -- & -- & X 
\end{longtable}

\subsection{AI in Mobile Emergency Communication apps} 
Mobile applications are crucial in modern emergency communication and early warning systems, significantly enhancing disaster response and management efficiency \citep{shaik2024}. Notably, approximately 70\% of emergency calls to the European emergency number 112 originate from mobile devices, highlighting their pivotal role according to the European Emergency Number Association (EENA) \citep{halliwell2018}. EENA stresses the importance of standardized mobile applications across the European Union \citep{repanovici2022}. These applications leverage smartphone technology and Global Navigation Satellite System (GNSS) chipsets to provide real-time location data and critical user information, facilitating swift rescue operations \citep{shaik2024}. They are especially beneficial for individuals, such as the deaf, who rely on them for emergency communication.

Advanced features like AI-driven chatbots and voice recognition significantly enhance user interaction with emergency apps \citep{zhu2022}. Voice recognition enables hands-free operation, which is crucial when physical input is impractical during emergencies. Chatbots, powered by AI, provide instant responses, guidance, and essential information to users \citep{adamopoulou2020}. AI also allows for personalized user profiles based on behavior and vulnerabilities, tailoring emergency communications and recommendations \citep{shaik2024}. For instance, AI-driven behavioral analysis predicts user actions and assesses vulnerabilities to deliver targeted assistance. Integration of AI extends to enhancing data visualization and situational awareness within emergency apps \citep{national2007}. AI can process diverse data sources such as weather forecasts, sensor networks monitoring environmental conditions, and social media for real-time updates \citep{tan2017}. Integrating multi-modular data provides comprehensive situational awareness, predicting hazards like hurricanes or earthquakes and pinpointing areas needing immediate aid.
Furthermore, AI automates alert systems based on predefined parameters, ensuring timely and accurate notifications without human intervention \citep{sarker2022}. Automation minimizes communication delays and enhances emergency response efficiency \citep{bayrak2009}. This multi-faceted AI integration ensures emergency apps remain responsive and effective in evolving disaster scenarios. In summary, mobile emergency applications enhanced with AI technology are vital in modern disaster management, offering real-time communication, personalized assistance, and comprehensive situational awareness to improve emergency response outcomes.

\subsection{AI in Risk Perception}
Often the chaotic nature of the emergency renders the possibility of satisfying public demand for timely information about the damage, mitigation actions, and active participation difficult. It is the role of an emergency communication specialist to communicate the scope and extent of the risk in a way that takes into account possible public adversity and ensures public trust and support. Especially, there is a gap between actual risk and perceived risk by the public. Specifically, risk perception is composed of two components: Hazard (data and facts about the threat and damage) and Outrage (how the threat makes the individual feel) \citep{Sandman1989}
\[Perceived Risk  =  Hazard  + Outrage\]
\begin{figure}
    \centering
    \includegraphics[width=0.95\linewidth]{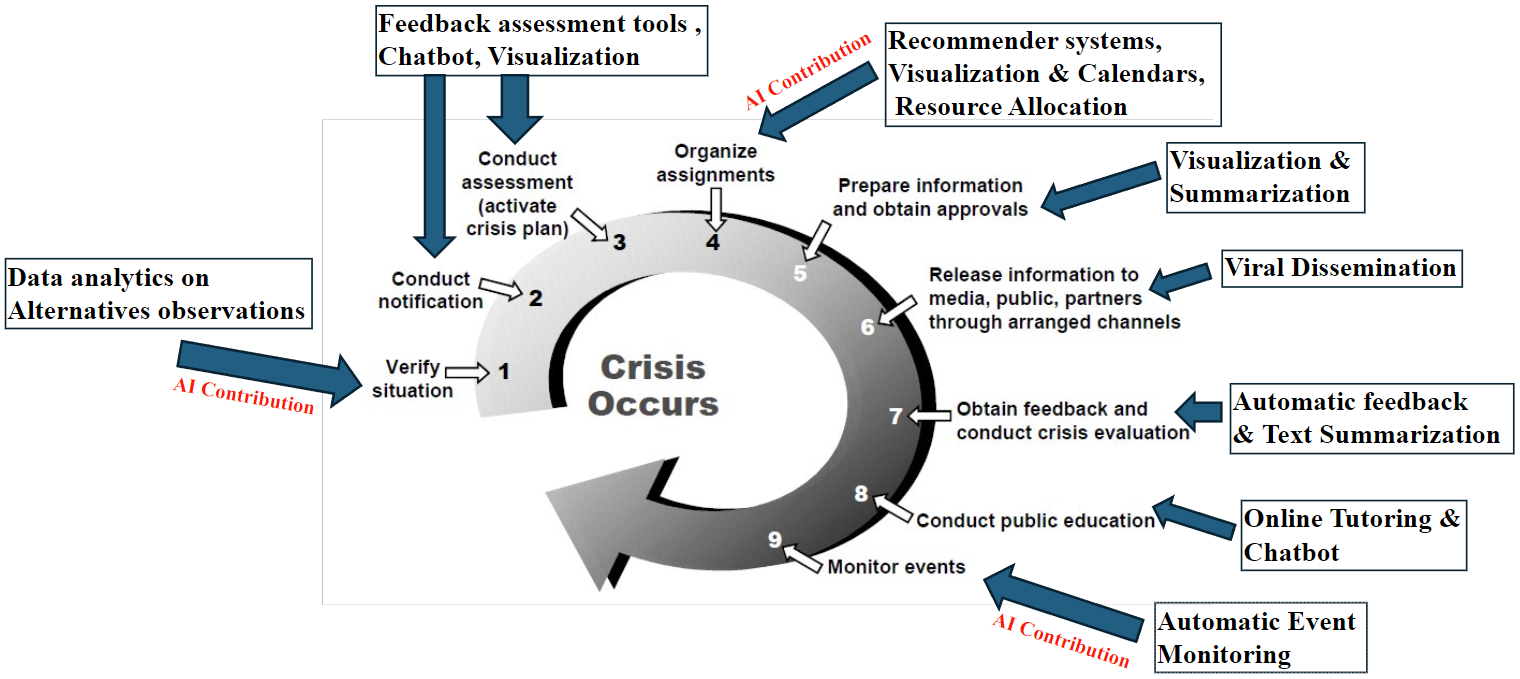}
    \caption{An illustration of AI in Reynolds's Crisis Response steps.}
    \label{fig:info1}
\end{figure}
People often react according to the strength of their outrage, which requires communication about the hazard to take into account public emotions. Common emotional responses include Anger, Fear, Depression, Empathy, and Anxiety. These emotions are further shaped by media coverage, damage at a personal and family level, personality, uncertainty surrounding the extent of the damage, and future outlook. For instance, the 24-hour-a-day coverage following the Oklahoma City bombing gave local children the impression that the entire town had been demolished, whereas, only a small part of it was affected. Catastrophic and immoral events often create long-lasting mental health effects that lead to anger, fear, and frustration. Psychological studies have shown that these emotions may cause problematic responses to emergency calls, such as \textit{Denial} (refusal to take good advice), \textit{Stigmatization} (isolation of individual or group ), or \textit{Panic} (irrational behavior) \citep{Cohn1990}. To take into account emotional state, Reynolds \cite{Reynolds2002} proposed a nine-step crisis response, which is widely acknowledged by practitioners, starting from fact verification, notification with various stakeholder groups, crisis assessment, organizational management, seeking various required approvals, release information to the public, conduct feedback and crisis evaluation, public education, and finally event monitoring.
In this respect, Fig. \ref{fig:info1} illustrates how various AI-based modules can be applied to Reynolds's nine steps of crisis response.

\label{sec:5}
\section{\MakeUppercase{Nordic countries Case Studies}}
The Nordic countries offer valuable insights into the application of AI in Early Warning Systems (EWS). Case studies from Finland, Denmark, and Norway exemplify how AI technologies have revolutionized emergency handling and monitoring, underscoring their potential to enhance disaster management across the region amidst climate change challenges. Norway faces diverse landslide risks due to its challenging terrain. In response, the Norwegian Water Resources and Energy Directorate (NVE) implemented a national Landslide Early Warning System (LEWS) in 2013. This system employs empirical models, real-time data, and expert judgment to issue daily warnings tailored to each municipality. Based on thresholds derived from historical landslide events, LEWS categorizes warnings into four levels (Green, Yellow, Orange, and Red) to indicate varying probabilities of landslide occurrence \citep{piciullo2016}. LEWS integrates diverse data sources including Hydrometeorological Index, Landslide Susceptibility Maps, and Real-Time Observations, to continuously assess landslide risks. It disseminates warnings via web tools like xgeo\footnote{\url{www.xgeo.no}}, regObs\footnote{\url{www.regObs.no}}, and varsom\footnote{\url{www.varsom.no}}, facilitating public and expert access to critical information. This approach has significantly improved preparedness and response capabilities, ensuring timely actions and enhancing public safety during landslide events \citep{luca2017}. In Denmark, flood forecasting in urban areas like Greve utilizes advanced AI techniques such as Deep Convolutional Generative Adversarial Networks (DCGAN). This real-time predictive model accurately forecasts flood propagation, aiding in early warning and urban flood management. The DCGAN's ability to maintain spatial and temporal accuracy over extended forecasting horizons supports effective flood hazard mitigation and water resource management strategies \citep{cheng2021, yao2024}.

Finland's "\textit{Luonnononnettomuuksien Varoitusjärjestelmä} (LUOVA)" (translates to Natural Disaster Warning System in English) \citep{luova2005} system exemplifies effective disaster communication and early warning integration. Managed by the Finnish Meteorological Institute (FMI), LUOVA consolidates forecasts, evaluations, and warnings from multiple sources to provide real-time situational awareness for authorities \citep{henriksen2018}. It supports preparedness and response efforts by disseminating alerts through various channels, including stationary sirens, broadcast media, and multilingual notifications. LUOVA's comprehensive approach ensures enhanced coordination, public safety, and improved decision-making during natural disasters \citep{henriksen2018}. These implementations highlight how AI-driven technologies enhance EWS in the Nordic countries. By integrating advanced modeling, real-time data analytics, and expert systems, these systems improve disaster prediction accuracy and response efficiency. Continuous innovation and adaptation are crucial in mitigating the impacts of climate change and natural disasters, making these technologies indispensable for effective disaster management strategies. To conclude, AI technologies are pivotal in transforming emergency management across the Nordic region, demonstrating their potential to safeguard communities and infrastructure from increasingly complex and frequent natural hazards. A summarized version of each country's case study can be observed in Table \ref{tabnordic}.

\begin{longtable}{>{\RaggedRight}p{1.5cm}>{\RaggedRight}p{3.5cm}>{\RaggedRight}p{5.5cm}>{\RaggedRight\arraybackslash}p{3.5cm}}
  \caption{Nordic Country EWS Case Studies} \label{tabnordic} \\
  \toprule
  EWS & Application & Impact & Characteristics\\
  \midrule
  \endfirsthead
  \multicolumn{4}{c}{{\tablename\ \thetable{} -- Continued from previous page}} \\
  \toprule
  EWS & Application & Impact & Characteristics\\
  \midrule
  \endhead
  \midrule
  \multicolumn{4}{r}{{Continued on next page}} \\
  \bottomrule
  \endfoot
  \bottomrule
  \endlastfoot
  LEWS & Build empirical models based on real-time data to achieve expert judgment to issue daily municipality-specific landslide warnings &  LEWS assesses landslide risks using diverse data, improving preparedness and response via timely public notifications & Uses the data from Hydrometeorological Index, Landslide Susceptibility Maps, and Real-Time Observations, to continuously assess landslide risks \\
  \\
  Flood forecasting & A DCGAN-based real-time predictive model accurately forecasts floods, enhancing early warning and urban management & The DCGAN maintains spatial-temporal accuracy, aiding in effective flood mitigation and water resource management & A Generator and Discriminator model architecture is trained on spatiotemporal datasets in which CNN layers are used to unravel the hidden information from the flow patterns \\
  \\
  LUOVA & LUOVA aggregates forecasts, evaluations, and warnings from various sources to deliver real-time situational awareness for authorities &  Integrating advanced modeling, real-time analytics, and expert systems enhances disaster prediction accuracy and response efficiency & The Finnish Meteorological Institute, the University of Helsinki's Institute of Seismology, and the Finnish Environment Institute are the sources of data for LUOVA\\
\end{longtable}

Furthermore, the role of mobile emergency applications in EWS in Nordic countries is considered crucial during crises or disasters \citep{shaik2024}. Applications such as SOS Alarm\footnote{\url{www.sosalarm.se/}} and Krisinformation\footnote{\url{www.Krisinformation.se}} in Sweden, Suomi 112\footnote{\url{www.112.fi/en/112-suomi-application}} for Finland, and Hjelp 113\footnote{\url{www.norskluftambulanse.no/eng/hjelp113/}} developed under the Norwegian Air Ambulance Foundation actively notify their users about the events to create situational awareness. These countries are constantly evolving to provide better Emergency Response Systems by integrating AI-based decision-making and mapping. For instance, in Sweden, SOS Alarm and the Danish business Corti\footnote{\url{www.corti.ai/company/about-corti}} work together to use AI to generate COVID-19 hotspot maps that help with crisis management and infection spread tracking \citep{SOSAlarm2020}. This program, which is a component of SOS Alarm's innovation initiatives, improves society's crisis response by utilizing real-time data. To facilitate resource allocation and preventive treatments, the AI evaluates call data to show hotspots without disclosing the precise locations of patients. The project intends to improve crisis prevention and analytical capacities of emergency services by extending AI applications to other domains, such as seasonal forest fires and influenza outbreaks \citep{SOSAlarm2020}. \label{sec:6}
\section{\MakeUppercase{Conclusion and Perspective Works}}
Globally, the frequency, intensity, and severity of natural catastrophes have all increased dramatically due to climate change. There has been a noticeable increase in the frequency of extreme weather phenomena such as heatwaves, droughts, floods, and storms. For instance, the World Meteorological Organization (WMO) reports a fivefold rise in weather-related disasters during the last 50 years \citep{rajabi2023}. The world's temperature has risen, with 2023 ranking among the warmest years ever recorded \citep{van2024}. Heatwaves have become increasingly frequent and severe, affecting millions of people and causing serious health problems and deaths. Rising sea levels brought on by melting glaciers and ice caps worsen coastal flooding. Low-lying locations are particularly risky, as some experience more regular and severe tidal flooding and storm surges. According to climate models, heatwave frequency and intensity are expected to rise significantly. Regions that now only see extreme heatwaves once every 20 years may see them every year by the end of the 21st century \citep{yin2022}. The Intergovernmental Panel on Climate Change (IPCC) predicts that heavy precipitation events will become more frequent and intense, increasing the likelihood of severe flooding in many parts of the world \citep{meresa2022}. Sea levels are expected to rise further; under high emission scenarios, projections imply an increase of up to 1 meter by 2100 \citep{frederikse2020}. This will majorly affect coastal areas, raising the possibility of erosion, flooding, and habitat loss. Forecasts for the future show that these patterns will continue, with longer droughts, flooding, tropical cyclones, and more frequent and severe heatwaves. Addressing these challenges requires urgent and sustained efforts in climate mitigation and adaptation to reduce risks and enhance resilience. 

The evolution of emergency communication protocols has had a profound impact on the development and application of AI-based systems in EWS. Modern communication protocols facilitate better data sharing between different agencies and organizations. This enhanced data sharing is crucial for AI systems, which rely on large datasets to train models and improve prediction accuracy \citep{kim2023}. For example, satellite imagery and remote sensing data shared across platforms can be integrated into AI models for more accurate hazard predictions. Advanced communication protocols enable real-time data transmission, which is essential for timely updates in AI models. Real-time data allows for continuous learning and adjustment of AI systems, leading to more precise and current early warning outputs. Enhanced communication protocols improve collaboration and coordination among various stakeholders, including international organizations, local governments, and communities. This collaboration ensures that AI-generated insights are effectively disseminated and acted upon. It also helps in bridging the "warning-response gap" by ensuring that warnings lead to timely and adequate responses. Apart from its pros, integrating AI technologies into EWS presents several challenges, particularly concerning data privacy, algorithmic bias, technical complexity, operational reliability, and ethical considerations. The European Commission's recent AI legislation further complicates this integration by imposing strict requirements, especially for high-risk applications like EWS. These include rigorous data governance, transparency, explainability, human oversight, and compliance with certification standards. Addressing these constraints necessitates a collaborative effort among policymakers, technologists, and stakeholders to ensure AI-driven EWS are both effective and trustworthy, ultimately enhancing disaster preparedness and response.\label{sec:7}
\section*{Acknowledgement}
This work is supported by the EU Chanse programme DiGEMERGE on emergency communications and the Finnish Digital Water (DIWA) flagship programme. 

\bibliographystyle{apalike}
\bibliography{reference}

\begin{thebibliography}{}
\bibitem[Smith, 2017]{smith2017ai}
Smith, B. (2017).
\newblock Ai for earth can be a game-changer for our planet.
\newblock {\em Retrieved from ai-for-earth-can-be-a-game-changer-for-our-planet/(Accessed September 12, 2022)}.
\end{thebibliography}

\end{document}